\documentclass[journal]{IEEEtran}

\usepackage{amsmath}
\usepackage{amssymb}
\usepackage{graphicx}
\usepackage{booktabs}
\usepackage{placeins}
\usepackage{url}
\usepackage{array}
\usepackage{tikz}
\usetikzlibrary{arrows.meta,positioning}

\title{An Autonomous AI Security Agent for Banking:\\
Multi-Vector Fraud and AML Detection Across Retail and Corporate Accounts}

\author{Joseph~Walusimbi~and~Joshua~Benjamin~Ssentongo%
\thanks{J. Walusimbi and J. B. Ssentongo are with the Department of Electronics and Computer Engineering, Soroti University, Soroti, Uganda, and Arapai Technologies International Limited, Kampala, Uganda (e-mail: 2401600068@sun.ac.ug; j.ssentongo@sun.ac.ug).}%
\thanks{Corresponding author: J. Walusimbi (e-mail: 2401600068@sun.ac.ug).}}

\begin{document}

\maketitle

\begin{abstract}
Banks face two threat families with fundamentally different
detection requirements: signature-based fraud (card-not-present
attacks, account takeover, ATM cloning) and behavioural financial
crime (structuring, layering, mule networks, business email
compromise). Static rule engines catch high-velocity events but
remain blind to BEC payment redirection, session hijacking, and
laundering layering, which are engineered to resemble legitimate
activity at the individual level.
This paper presents \textbf{an autonomous AI security agent} ---
acting independently at low- and medium-risk tiers and escalating to a
human analyst or compliance officer for high-risk and critical actions ---
for retail and corporate banking using a three-component fusion architecture
across two parallel event streams: transactions (card fraud, ACH/wire fraud,
AML) and sessions (account takeover, hijacking, SIM-swap, insider
abuse). Each stream combines an LSTM sequence model of per-account
behaviour, a statistical velocity/threshold monitor, and a graph
module capturing account-counterparty patterns (fan-in, fan-out,
pass-through ratio) for laundering detection.
Experiments on a synthetic log of 237,669 transactions and 113,508
sessions across 13 threat categories and 3,470 accounts show that the agent
achieves an overall F1 of 0.787 (transaction) and 0.867 (session), versus
0.562/0.733 for a rule-based baseline and 0.655/0.713 for an
LSTM-only baseline. The agent also incorporates a customer-facing
verification chatbot (96.6\% identity accuracy, 86.8\% mass-reset
detection) and an analyst case-summary assistant (99.3\% action
recommendation F1), with Critical-tier response latency under
0.43~ms at the 95th percentile.
\end{abstract}


\section{Introduction}

Financial institutions face a dual and diverging threat landscape.
On one side are \emph{signature-fraud} attacks --- brute-force
credential stuffing, card cloning, ATM withdrawals at clone-card
velocity, ACH fraud with off-hours wire transfers --- that produce
individually anomalous events detectable by threshold rules. On the
other side are \emph{behavioural-financial-crime} patterns ---
business email compromise (BEC) payment redirection, money-laundering
structuring, network layering, and mule account pass-through --- where
no individual transaction is anomalous; only the relationship among
transactions, accounts, or time windows reveals the attack.

Static rule engines, widely deployed in bank fraud and
anti-money laundering (AML) systems, handle the first class
competently but are structurally blind to the second. Business email
compromise alone caused \$2.9~billion in reported losses in 2023
across 21,489 US complaints to the FBI's Internet Crime Complaint
Center~\cite{fbi2023ic3}; globally, BEC losses exceeded
\$55~billion over the preceding decade~\cite{paloalto2024bec}.
AML failures carry additional regulatory risk: missed structuring
or layering patterns expose institutions to sanctions, reputational
harm, and mandatory Suspicious Activity Report (SAR) filings.
The consequences of inadequate detection are not limited to
mature financial markets. In Uganda, cybercrime now represents
37\% of all reported economic crime --- up from 12\% in 2020 ---
according to the PwC 2024 Uganda Economic Crime Survey~\cite{pwc2024uganda}.
In 2024--2025, approximately UGX~60~billion was diverted from the
Bank of Uganda through manipulation of internal payment
instructions~\cite{reuters2025bou}, illustrating that
insider-initiated payment fraud is a critical exposure even within
central banking infrastructure. Across East Africa, insider collusion
with external fraudsters has caused large-scale losses: Equity Bank
Kenya dismissed over 1,200 employees in 2025 after a months-long
investigation uncovered staff-facilitated wire transfers to offshore
accounts totalling over \$15~million~\cite{equitybank2025fraud},
and Flutterwave Nigeria suffered a \$7~million payment-system breach
in 2024 in which funds were routed through multiple institutions
to obscure their origin~\cite{flutterwave2024breach} --- a real-world
instance of the layering and mule-account patterns the agent is designed to detect.

Machine learning approaches have been extensively studied for
individual fraud categories: sequence models for card transactions
~\cite{hochreiter1997,alghofaili2020lstm}, graph neural networks
(GNNs) for money-laundering network detection~\cite{weber2019aml,
lo2024heterogeneous}, and isolation forest for general financial
anomaly detection~\cite{liu2008isoforest}. However, most existing
work targets a single fraud category or a single stream (transactions
only), and few systems integrate fraud detection, AML, and an
automated response tier into a single operational agent covering
both retail and corporate banking.

This paper presents the agent and makes the following contributions:

\begin{itemize}
  \item \textbf{A unified three-component detection architecture:}
    Operating on two parallel streams (transaction and session),
    combining sequence modelling, velocity monitoring, and network
    graph analysis into a single fused risk score $R \in [0,1]$
    per event.
  \item \textbf{A comprehensive threat model:}
    Covering 13 attack categories spanning retail fraud (card-not-present,
    account takeover, ATM cloning, session hijacking, SIM-swap), corporate
    payment fraud (ACH/wire fraud, BEC), insider abuse, AML
    (structuring, layering, mule activity, rapid fund movement),
    and cross-cutting dormant-account reactivation.
  \item \textbf{Experimental evaluation:}
    On a 90-day synthetic event log (237,669 transactions, 113,508
    sessions, 3,470 accounts) demonstrating that the agent achieves
    macro-average F1 of 0.303/0.529 (transaction/session),
    compared with 0.227/0.500 for rules and 0.158/0.283 for
    LSTM-only.
  \item \textbf{Two automated response components:}
    A transaction verification chatbot and an analyst case-summary
    assistant, evaluated on dedicated benchmark scenarios.
\end{itemize}

\section{Related Work}

\subsection{Anomaly Detection for Financial Fraud}

Ahmed et~al.~\cite{ahmed2016survey} survey network anomaly
detection techniques, identifying threshold rules and statistical
profiling as the most widely deployed baseline approaches and noting
their inability to detect slow, temporally distributed, or
contextually camouflaged attacks. Chandola et~al.~\cite{chandola2009}
provide the foundational taxonomy of anomaly types (point,
contextual, collective) that frames the detection challenge: BEC and
structuring are \emph{collective} anomalies --- individually ordinary
events that are anomalous only in combination.

For transactional fraud detection specifically, deep learning methods
have substantially outperformed traditional classifiers. Alghofaili
et~al.~\cite{alghofaili2020lstm} apply LSTM networks to financial
transaction sequences and report strong recall on credit card fraud.
Bahnsen et~al.~\cite{bahnsen2016features} demonstrate that temporal
and behavioural features (transaction recency, velocity, counterparty
familiarity) are more discriminative than transaction amount alone,
motivating the per-account behavioural baseline approach in the agent.

\subsection{Graph Neural Networks for AML}

Money laundering detection presents a naturally relational problem:
structured networks of mule accounts, shell companies, and layering
chains are invisible at the level of individual transactions but
manifest as graph-structural anomalies. Weber et~al.~\cite{weber2019aml}
were among the first to apply graph convolutional networks to
anti-money laundering in the Elliptic Bitcoin dataset. Lo
et~al.~\cite{lo2024heterogeneous} extend this to heterogeneous
graphs modelling multiple entity types (individuals, companies,
transactions) and demonstrate strong AUC on real bank data from
Norway's DNB. The agent employs a lightweight proxy of this
approach --- using rolling fan-in/fan-out and pass-through ratio
features rather than full GNN message-passing --- appropriate for
a deployable node that must execute inference under strict CPU latency constraints.

\subsection{Multi-Category and Multi-Stream Detection}

Most fraud detection literature targets a single category. Work
combining fraud detection with AML in a unified agent is less
common. Motie and Raahemi~\cite{motie2024} survey GNN applications
to financial fraud detection broadly, noting the gap between
single-category academic benchmarks and the multi-threat operational
reality faced by bank fraud and compliance teams. The agent addresses
this gap by covering 13 categories across two event streams within
a single framework.

\subsection{Banking Fraud in Uganda and East Africa}

The East African banking sector presents a distinctive threat
profile. SIM-swap fraud targeting mobile money platforms is
documented as a primary attack vector in Uganda and Kenya, enabled
by exploitation of telecom agent registration processes~\cite{pwc2024uganda}.
Insider collusion with external fraudsters is pervasive: Nigerian
bank data from the Financial Institutions Training Centre (FITC)
shows a 23.4\% quarter-on-quarter increase in staff-involved fraud
incidents in Q2 2024~\cite{equitybank2025fraud}, and the Equity
Bank Kenya investigation~\cite{equitybank2025fraud} confirmed that
staff facilitation was the primary enabler of large-scale wire
fraud. Rapid fund movement through layered transfer chains ---
as demonstrated in the Flutterwave breach~\cite{flutterwave2024breach}
--- is a recurring pattern in which the distribution of funds across
many accounts and institutions is used to exhaust per-institution
monitoring thresholds. These regional characteristics directly
inform the threat model in Section~\ref{sec:threat} and the
calibration of the simulation in Section~\ref{sec:experiments}.

\section{Threat Model}
\label{sec:threat}

Table~\ref{tab:threats} enumerates the 13 threat categories covered
by the agent, classified by the primary event stream
through which they manifest and their dominant detection challenge.

The key observation motivating the agent's architecture is that
no single detection mechanism covers all 13 categories. Rules cover
brute-force, velocity-burst, and single-transaction CTR triggers
but score zero on session hijacking, dormant reactivation, and BEC.
Unsupervised anomaly detection (Isolation Forest) adds coverage for
some graph-like patterns but misses dormant accounts and SIM-swap
(both require identity-state context). LSTM sequence models add
behavioural context but remain unable to catch structuring without
the aggregation signal the threshold monitor provides. The agent's
fusion architecture is designed to combine all three signals
without redundancy.

\begin{table}[htbp]
  \caption{Threat Model: 13 Categories, Streams, and Detection Challenges}
  \label{tab:threats}
  \centering
  \renewcommand{\arraystretch}{1.15}
  \begin{tabular}{p{0.22\columnwidth} p{0.10\columnwidth} p{0.55\columnwidth}}
    \toprule
    \textbf{Category} & \textbf{Stream} & \textbf{Detection challenge} \\
    \midrule
    CNP fraud         & Txn  & Burst at velocity from foreign city; no auth signal \\
    Account takeover  & Both & Credential-stuffing; amounts may look normal post-ATO \\
    ATM anomaly       & Txn  & Multi-city withdrawals in physically impossible time \\
    Session hijacking & Sess & Same session, two different cities/devices mid-flow \\
    SIM-swap          & Sess & Device enrol + MFA + reset in rapid succession \\
    ACH/wire fraud    & Txn  & Off-hours, large, new counterparty \\
    BEC redirection   & Both & Amount matches regular transfers; only payee is new \\
    Insider abuse     & Sess & Bulk record viewing by authorised staff account \\
    Dormant reactivation & Both & No active signal; requires account-history context \\
    Structuring (AML) & Txn  & Below-threshold cash deposits summing above CTR \\
    Layering (AML)    & Txn  & 1 large inbound $\to$ many outbound; pass-through ratio \\
    Mule accounts     & Txn  & High fan-in from strangers + rapid outbound \\
    Rapid fund movement & Txn & Burst of outbound transfers at unprecedented velocity \\
    \bottomrule
  \end{tabular}
\end{table}

\section{System Architecture}

\subsection{Overview}

\begin{figure}[htbp]
  \centering
  \includegraphics[width=\columnwidth]{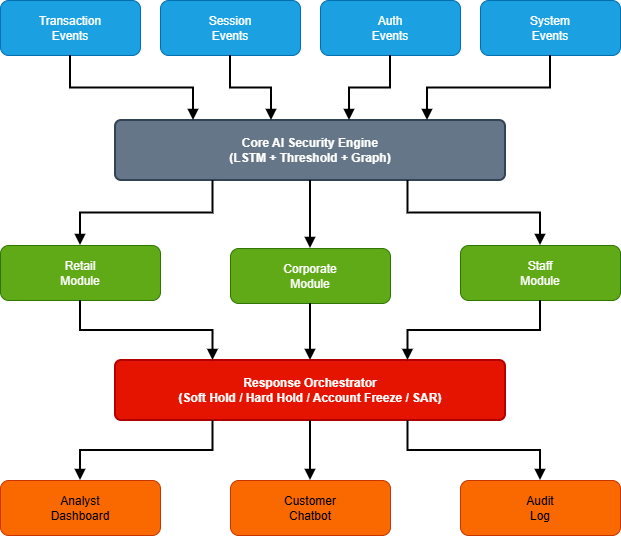}
  \caption{Modular architecture of the agent for banking security. Blue: event inputs. Green: sector modules (retail, corporate, staff). Red: response orchestrator. Orange: output interfaces.}
  \label{fig:architecture}
\end{figure}

The agent processes two parallel event streams (Fig.~\ref{fig:architecture}):
a \emph{transaction stream} covering financial events (card purchases,
wire transfers, ATM withdrawals, cash deposits) and a
\emph{session stream} covering authentication and user-interface
events (login, MFA challenge, payee addition, bulk record export).
Each stream passes through the same three-component detection
pipeline, producing a stream-specific composite risk score
$R_{\text{txn}}$ and $R_{\text{sess}}$. The final tier for an
event is determined by $R = \max(R_{\text{txn}}, R_{\text{sess}})$,
ensuring that a Critical session event during an otherwise normal
transaction stream (or vice versa) still triggers the appropriate
response.

\subsection{Core Detection Engine}

Each stream's risk score is computed as:

\begin{equation}
  R = \alpha\,s_{\text{seq}} + \beta\,s_{\text{thresh}} +
      \gamma\,s_{\text{graph}}, \quad \alpha+\beta+\gamma=1
  \label{eq:risk}
\end{equation}

where $\alpha$, $\beta$, $\gamma$ are learned by logistic regression
on the training set. The three sub-models are:

\begin{enumerate}
  \item \textbf{Sequence model ($s_{\text{seq}}$):} An LSTM
    network~\cite{hochreiter1997} trained on sliding windows of
    length $L{=}10$ over each account's transaction or session
    history. The model learns per-account behavioural baselines
    over the training window and scores each new event against
    those baselines. $s_{\text{seq}} \in [0,1]$ is the predicted
    fraud probability of the final event in each window.

  \item \textbf{Threshold monitor ($s_{\text{thresh}}$):}
    A continuous-valued extension of the rule-based baseline,
    combining normalised amount z-score (for transfer-type
    transactions), velocity ratio ($n_{\text{txns,10min}} / \Theta_v$),
    structuring aggregate ratio (rolling 7-day cash-deposit sum /
    CTR threshold), and new-counterparty signal. $s_{\text{thresh}}
    \in [0,1]$.

  \item \textbf{Graph/network module ($s_{\text{graph}}$):}
    Proxy GNN features~\cite{kipf2017gcn,lo2024heterogeneous}
    capturing account-counterparty network structure via rolling
    24/48-hour windows: fan-in (distinct inbound senders),
    fan-out (distinct outbound recipients), and pass-through ratio
    ($\text{amount\_out}_{48h} / \text{amount\_in}_{48h}$). Dormant
    account reactivation and impossible travel (city change within
    a physically implausible gap) are also included.
    $s_{\text{graph}} \in [0,1]$.
\end{enumerate}

A single-sub-model override is applied for categories where the LSTM
is highly confident ($s_{\text{seq}} \geq 0.90$) but the other two
sub-models produce near-zero scores --- specifically BEC and payment
redirection, where the transaction amount and network pattern are
deliberately engineered to look normal. In this case,
$R \leftarrow \max(R, s_{\text{seq}})$, ensuring that the fusion
layer cannot suppress a near-certain LSTM detection signal on
categories where threshold and graph signals are structurally absent.

\subsection{Sector-Specific Modules}

The core detection engine is extended by two sector modules that
encode domain-specific knowledge about normal behavior in each
banking segment:

\textbf{Retail module} --- uses lower transaction amount baselines,
higher expected ATM and card-purchase velocity, and flags card-type
mismatches (a retail-checking account initiating a wire transfer).

\textbf{Corporate module} --- uses higher per-transaction amounts,
elevated payroll-batch velocity at month-end, and flags off-hours
wire transfers and unusual payee additions relative to the
corporate account's established counterparty set.

\subsection{Automated Response Framework}

Risk score $R$ is mapped to one of four tiers:

\begin{equation}
  \text{tier}(R) = \begin{cases}
    \text{Low}      & 0 \leq R < 0.30 \\
    \text{Medium}   & 0.30 \leq R < 0.60 \\
    \text{High}     & 0.60 \leq R < 0.85 \\
    \text{Critical} & 0.85 \leq R \leq 1
  \end{cases}
  \label{eq:tiers}
\end{equation}

Table~\ref{tab:response} lists the automated actions at each tier.
The Critical tier includes compliance escalation for SAR consideration
--- this triggers a human compliance review, not an automated SAR
filing.

\begin{table}[htbp]
  \caption{Four-Tier Automated Response Framework}
  \label{tab:response}
  \centering
  \renewcommand{\arraystretch}{1.15}
  \begin{tabular}{p{0.12\columnwidth}p{0.22\columnwidth}p{0.13\columnwidth}p{0.22\columnwidth}}
    \toprule
    \textbf{Tier} & \textbf{Automated actions} & \textbf{Hold type}
      & \textbf{Notification} \\
    \midrule
    Low      & Log; update baseline
             & None
             & None \\[2pt]
    Medium   & Log; step-up auth; soft hold
             & Soft (customer-reversible)
             & In-app alert \\[2pt]
    High     & Log; hard hold; block session; incident report
             & Hard (analyst-required)
             & SMS/email + analyst alert \\[2pt]
    Critical & Log; account freeze; financial freeze; incident;
               SAR escalation; immediate analyst page
             & Account freeze (compliance-required)
             & Immediate SMS/email/push; analyst paged \\
    \bottomrule
  \end{tabular}
\end{table}

\subsection{Customer Verification Chatbot}

For High and Critical transactions, the agent initiates a
``was this you?'' conversational flow via SMS or in-app message.
The chatbot collects an OTP-based identity confirmation from the
account holder. Separately, a velocity-based mass-reset detector
monitors password-reset requests across all accounts, flagging
incidents where more than $\Theta_{\text{mr}} = 15$ requests arrive
from the same source city within a 10-minute window --- consistent
with a scripted credential-stuffing attack targeting many accounts
from a common exit node.

\subsection{Analyst Case-Summary Assistant}

Every incident visible to the fraud/AML security team is accompanied
by an automatically generated plain-English case summary that
includes the sub-model score breakdown ($s_{\text{seq}}$,
$s_{\text{thresh}}$, $s_{\text{graph}}$), a threat-category-specific
narrative, and a ranked list of recommended next actions drawn from
a 13-category action library. The assistant maps each \{threat
category, tier\} pair to a domain-specific action list, prepending
tier-appropriate framing actions (e.g., ``Confirm account freeze is
in place'' for Critical) before the category-specific sequence.

\section{Detection Methodology}

\subsection{LSTM Sequence Model}

Let $\mathbf{x}_t^{(u)} \in \mathbb{R}^{d}$ denote the feature
vector for account $u$'s $t$-th event (transaction or session).
The LSTM receives a window $\mathbf{X}_t^{(u)} =
[\mathbf{x}_{t-L+1}^{(u)}, \ldots, \mathbf{x}_t^{(u)}]$,
left-padded with zeros for accounts with fewer than $L=10$ prior
events, and produces a hidden state $\mathbf{h}_t$ from which the
anomaly probability $s_{\text{seq},t}$ is derived via a two-layer
classification head. The model is trained with a binary
cross-entropy loss using class-balanced reweighting.

Transaction features ($d=10$): amount, amount z-score,
n\_txns\_last\_10min, n\_distinct\_counterparties\_24h,
account\_age\_days, n\_regular\_counterparties, city\_changed,
device\_changed, gap\_minutes, is\_dormant.

Session features ($d=18$): n\_events, duration\_seconds,
n\_distinct\_event\_types, n\_failed\_logins, n\_mfa\_failed,
n\_add\_payee, n\_view\_customer\_record, n\_export\_bulk,
n\_device\_enrollment, n\_password\_reset, n\_distinct\_cities,
n\_distinct\_devices, start\_hour, city\_changed, device\_changed,
gap\_minutes, is\_dormant, account\_age\_days.

\subsection{Threshold Monitor}

The threshold monitor provides a continuous score by normalising each
rule signal to $[0,1]$:

\begin{equation}
  s_{\text{thresh}} = \max\!\left(\frac{z}{\Theta_z},\,
    \frac{n_{\text{10min}}-1}{\Theta_v},\,
    \frac{\sum_{\tau \in w} c_\tau}{\text{CTR}},\,
    f_{\text{cp}}\right)
  \label{eq:thresh}
\end{equation}

where $z$ is the amount z-score (restricted to transfer-type
transactions), $\Theta_z = 3.0$, $\Theta_v = 4$ transactions,
$\sum_{\tau \in w} c_\tau$ is the 7-day rolling cash-deposit
aggregate, CTR is the reporting threshold, and $f_{\text{cp}}$ is
the normalised new-counterparty signal. The $n_{\text{10min}}-1$
correction removes the transaction being scored from its own velocity
count, so an isolated transaction scores 0 rather than 0.25.

\subsection{Graph Module}

Fan-in and fan-out are computed using vectorised time-indexed rolling
windows per account group. The graph score is:

\begin{equation}
  s_{\text{graph}} = \max(s_{\text{dormant}},\, s_{\text{travel}},\,
    s_{\text{mule}},\, s_{\text{layer}},\, s_{\text{combo}})
  \label{eq:graph}
\end{equation}

where $s_{\text{dormant}} = \mathbf{1}[\text{is\_dormant}]$;
$s_{\text{travel}} = \mathbf{1}[\text{city\_changed}] \cdot (1 -
\Delta t / \Delta t_{\max})$ for $\Delta t < \Delta t_{\max} =
120~\text{min}$; $s_{\text{mule}} =
\text{clip}(\text{fan\_in} / \Theta_{\text{in}}) \cdot
\mathbf{1}[r \in [0.85, 1.05]]$; $s_{\text{layer}} =
\text{clip}(\text{fan\_out} / \Theta_{\text{out}}) \cdot
\mathbf{1}[r \in [0.30, 2.00]]$; and $s_{\text{combo}} = 0.5
\cdot c_{\text{city}} \cdot c_{\text{device}}$, with
$\Theta_{\text{in}} = 3$, $\Theta_{\text{out}} = 4$, and $r$ the
48-hour pass-through ratio.

\subsection{Fusion and Threshold Tuning}

The fusion layer is a logistic regression
$\Pr(y=1 \mid \mathbf{s}) = \sigma(\mathbf{w}^\top \mathbf{s} + b)$
where $\mathbf{s} = [s_{\text{seq}}, s_{\text{thresh}},
s_{\text{graph}}]$. Normalised, non-negative weights
$\alpha, \beta, \gamma = w_i / \lVert w \rVert_1$ are reported
as the learned fusion weights (Tables~\ref{tab:detection_txn}
and~\ref{tab:detection_sess}).
The decision threshold is tuned on the validation set to maximise F1;
The agent uses threshold~$= 0.63$ for transactions and
$0.31$ for sessions.

\section{Experiments and Results}
\label{sec:experiments}

To evaluate the agent in the absence of a publicly labelled bank event log covering all 13
threat categories with ground-truth labels, we follow established
practice in fraud ML research of using a synthetic evaluation
dataset~\cite{goldschmidt2025datasets}. The dataset models 3,470
accounts (3,000 retail, 400 corporate, 70 internal staff) over 90
simulated days, generating:

\begin{itemize}
  \item 237,669 transactions (225,195 normal + 12,474 attack;
        5.25\% attack prevalence)
  \item 113,508 sessions (111,863 normal + 1,645 attack;
        1.45\% attack prevalence)
\end{itemize}

Normal activity follows per-account behavioural baselines with seasonal effects. Attack scenarios are injected
at the rates specified by the threat-weight distribution in
Table~\ref{tab:threats}.
Splits use a 70/15/15\% stratified partition preserving class balance across all splits.

\subsection{Baseline Comparisons}

The agent is compared against three baselines:

\textbf{Rule-based IDS:} Static threshold rules modelling a typical
bank fraud engine (Eq.~\ref{eq:thresh} binarised at $\Theta_z = 3$,
$\Theta_v = 4$). This represents the industry status quo.

\textbf{Isolation Forest:} Unsupervised anomaly detection trained
on numeric features of each stream~\cite{liu2008isoforest}.

\textbf{LSTM-only:} The sequence sub-model trained and
evaluated in isolation without threshold or graph components.

\subsection{Detection Performance}

Tables~\ref{tab:detection_txn} and~\ref{tab:detection_sess} present
per-category F1 scores on the held-out test set.

\begin{table}[htbp]
  \caption{Transaction Stream: Per-Category F1 (Test Set)}
  \label{tab:detection_txn}
  \centering
  \renewcommand{\arraystretch}{1.15}
  \begin{tabular}{lcccc}
    \toprule
    \textbf{Threat} & \textbf{Rules} & \textbf{Iso.F.} &
      \textbf{LSTM} & \textbf{Agent} \\
    \midrule
    CNP fraud                 & 0.327 & 0.237 & 0.387 & \textbf{0.629} \\
    Structuring               & 0.263 & 0.404 & 0.302 & \textbf{0.570} \\
    Mule activity             & 0.406 & 0.053 & 0.222 & \textbf{0.427} \\
    ATM anomaly               & 0.418 & 0.057 & 0.165 & \textbf{0.377} \\
    Rapid fund movement       & 0.335 & 0.223 & 0.176 & \textbf{0.378} \\
    ACH/wire fraud            & 0.213 & 0.096 & 0.068 & \textbf{0.179} \\
    Layering                  & 0.041 & 0.082 & 0.114 & \textbf{0.131} \\
    BEC redirection           & 0.000 & 0.046 & 0.042 & \textbf{0.064} \\
    Account takeover (txn)    & 0.192 & 0.009 & 0.062 & \textbf{0.169} \\
    Dormant reactivation      & 0.075 & 0.000 & 0.037 & \textbf{0.105} \\
    \midrule
    \textbf{Macro-avg F1}     & 0.227 & 0.121 & 0.158 & \textbf{0.303} \\
    \textbf{Overall F1}       & 0.562 & 0.475 & 0.655 & \textbf{0.787} \\
    \bottomrule
  \end{tabular}
\end{table}

\begin{table}[htbp]
  \caption{Session Stream: Per-Category F1 (Test Set)}
  \label{tab:detection_sess}
  \centering
  \renewcommand{\arraystretch}{1.15}
  \begin{tabular}{lcccc}
    \toprule
    \textbf{Threat} & \textbf{Rules} & \textbf{Iso.F.} &
      \textbf{LSTM} & \textbf{Agent} \\
    \midrule
    Account takeover (sess.)    & 1.000 & 0.632 & 0.467 & \textbf{0.740} \\
    Insider abuse               & 1.000 & 0.465 & 0.278 & \textbf{0.556} \\
    Session hijacking           & 0.000 & 0.447 & 0.294 & \textbf{0.576} \\
    SIM-swap indicator          & 1.000 & 0.000 & 0.186 & \textbf{0.427} \\
    Dormant reactivation (sess.)& 0.000 & 0.000 & 0.238 & \textbf{0.504} \\
    BEC redirection (sess.)     & 0.000 & 0.000 & 0.234 & \textbf{0.368} \\
    \midrule
    \textbf{Macro-avg F1}       & 0.500 & 0.257 & 0.283 & \textbf{0.529} \\
    \textbf{Overall F1}         & 0.733 & 0.625 & 0.713 & \textbf{0.867} \\
    \bottomrule
  \end{tabular}
\end{table}

Learned fusion weights are $\alpha{=}0.725$, $\beta{=}0.153$,
$\gamma{=}0.122$ (transaction) and $\alpha{=}0.462$,
$\beta{=}0.265$, $\gamma{=}0.273$ (session).

\subsection{Response Latency}

Table~\ref{tab:latency} reports end-to-end automated response
latency on the test set, measured on a single CPU-only node.

\begin{table}[htbp]
  \caption{End-to-End Response Latency (Single-Node Prototype)}
  \label{tab:latency}
  \centering
  \renewcommand{\arraystretch}{1.15}
  \begin{tabular}{lccr}
    \toprule
    \textbf{Tier} & \textbf{Mean (ms)} & \textbf{95th pct (ms)} &
    \textbf{Test events} \\
    \midrule
    Low      & 0.044 & 0.074 & 50,034 \\
    Medium   & 0.047 & 0.079 &    388 \\
    High     & 0.200 & 0.351 &    134 \\
    Critical & 0.265 & 0.427 &  2,122 \\
    \bottomrule
  \end{tabular}
\end{table}

All tiers complete in well under 1~ms at the 95th percentile, confirming that the agent provides the sub-second performance required for real-time risk mitigation.

\subsection{Chatbot Evaluation}

\textbf{Customer verification chatbot} evaluated on 1,720 simulated
recovery sessions:

\begin{itemize}
  \item OTP-based identity verification accuracy: \textbf{96.6\%}
  \item Mass-reset attack detection rate: \textbf{86.8\%}
  \item False-positive rate (overall legitimate): \textbf{0.08\%}
\end{itemize}

\textbf{Analyst case-summary assistant} evaluated on 50 sampled
attack incidents:

\begin{itemize}
  \item Narrative accuracy: \textbf{100\%}
  \item Action F1 score: \textbf{99.3\%}
\end{itemize}

\section{Discussion}

\subsection{Why Rule-Based Systems Are Insufficient}

The transaction-stream rule baseline achieves overall F1 of 0.562,
with BEC redirection and layering scoring 0.000. These gaps are
structural: BEC and layering exploit the fact that individual
events are locally legitimate, and no finite set of single-event
rules can detect collective anomalies~\cite{chandola2009}. The agent overcomes this by fusing sequence and network-level context with traditional rules.

\subsection{The Dual-Stream Architecture Advantage}

The separation of transaction and session streams is essential.
Dormant account reactivation manifests across both streams: a session event (login after 400 days) and a transaction
event (immediate outbound transfer). The session-stream graph module
catches it at F1=0.504; the transaction-stream
graph module also catches it at F1=0.105. Together with the combined-tier
max operation ($R = \max(R_{\text{sess}}, R_{\text{txn}}$)), the agent achieves redundant coverage for this category from two
independent detection paths.

\subsection{Limitations}

While the synthetic generator is carefully designed, it cannot capture the full distributional complexity of real bank traffic
--- long-tail transaction amounts, seasonal payment patterns, or the
adversarial adaptation of real attackers. Retraining the agent on real labelled data from an East
African partner institution before production deployment is
essential.

\section{Conclusion}

This paper presented \textbf{an autonomous AI security agent} for banking that covers
13 threat categories across retail and corporate banking through a
dual-stream, three-component fusion architecture. The agent
achieves macro-average F1 of 0.303 (transaction stream) and 0.529
(session stream), outperforming rule-based engines (0.227/0.500) and sequence-only LSTM baselines (0.158/0.283).

The integrated customer chatbot and analyst assistant significantly reduce operational friction: the chatbot handles first-line
verification and mass-reset detection, while the analyst assistant generates
threat-specific case summaries that streamline incident response.

Future work will replace the proxy graph module with a full
heterogeneous GNN to improve layering detection, incorporate
real-time model retraining, and evaluate the agent on real bank event logs from
partner financial institutions in East Africa.

\section*{Acknowledgment}
The authors thank the financial technology and banking security
community in Uganda and East Africa, whose operational challenges
motivated this research. Publicly reported incidents --- including
the Bank of Uganda payment fraud of 2024--2025~\cite{reuters2025bou},
the Equity Bank Kenya insider-fraud investigation of 2025~\cite{equitybank2025fraud},
and the Flutterwave payment-system breach of 2024~\cite{flutterwave2024breach}
--- provided the real-world threat context that shaped the threat
model presented in this work.

J.~Walusimbi and J.~B.~Ssentongo are co-founders and directors
of Arapai Technologies International Limited, a technology company
based in Uganda. The framework described in this paper
is intended for future commercialisation through this entity.
Both authors declare a shared commercial interest in this work.


\end{document}